\documentclass[twocolumn,showpacs,preprintnumbers,amsmath,amssymb]{revtex4}

\usepackage{graphicx}
\usepackage{dcolumn}
\usepackage{bm}

\begin{document}


\title{Temporal Distinguishability of an N-Photon State\\ and Its Characterization by Quantum Interference}

\author{Z. Y. Ou}
 \affiliation{Department of Physics, Indiana
University-Purdue University Indianapolis \\ 402 N. Blackford
Street, Indianapolis, IN 46202}

\date{\today}

\begin{abstract}
We present a multi-mode model to describe an arbitrary N-photon
state with a wide spectral range and some arbitrary temporal
distribution. In general, some of the $N$ photons are spread out
in time while other may overlap and become indistinguishable. From
this model, we find that the temporal (in)distinguishability of
photons is related to the exchange symmetry of the multi-photon
wave function. We find that simple multi-photon detection scheme
gives rise to a more general photon bunching effect with the
famous two-photon effect as a special case. We then send this
N-photon state into a recently discovered multi-photon
interference scheme. We calculate the visibility of the
multi-photon interference scheme and find that it is related to
the temporal distinguishability of the $N$ photons.  Maximum
visibility of one is achieved for the indistinguishable N-photon
state whereas the visibility degrades when some of the photons are
separated and become distinguishable. Thus we can identify an
experimentally measurable quantity that may quantitatively define
the degree of indistinguishability of an N-photon state. This
presents a quantitative demonstration of the complementary
principle of quantum interference.
\end{abstract}

\maketitle

\section{\label{sec:level1}Introduction}

The coherence properties of an optical field are best described by
the field correlation function in space and time \cite{mw}. Most
commonly used quantity to characterize the coherence property of
an optical field is the coherence time or coherence length for
temporal coherence. Roughly speaking, the coherence length of an
optical field is the distance within which the field can be
described as a single uninterrupted wave train. In other words,
any two points within the coherence length will have a fixed phase
relationship. However, this description is primarily concerned
with wave aspect of an optical field and is based on the
interference effect observed in intensity or single photon
interference effect. More specifically in terms of the quantum
coherence theory \cite{glau}, it is related to the field
correlation function of
\begin{eqnarray}
\Gamma(\tau) = \langle \hat E^{(-)}(t+\tau) \hat E^{(+)}(t)
\rangle,\label{1}
\end{eqnarray}
where
\begin{eqnarray}
\Big[\hat E^{(-)}\Big]^{\dag} = \hat E^{(+)}(t)
={1\over\sqrt{2\pi}}\int d\omega\hat a(\omega) e^{-i\omega
t}\label{E}
\end{eqnarray}
for a quasi-monochromatic field \cite{man66} and the average is
over the quantum state of the field. The visibility of the
single-photon interference fringes is simply the absolute value of
the normalized field correlation function \cite{bw}:
\begin{eqnarray}
\gamma(\tau) = \Gamma(\tau)/\Gamma(0). \label{2}
\end{eqnarray}
However, this description becomes rudimentary when we start to
deal with the cases involving more than one photon in quantum
information. One may use a higher order correlation function such
as the intensity correlation function \cite{glau}
\begin{eqnarray}
&&\Gamma^{(N)}(t_1,t_2,...,t_N) \cr &&\hskip 0.2in = \langle \hat
E^{(-)}(t_1)...\hat E^{(-)}(t_N) \hat E^{(+)}(t_N)...\hat
E^{(+)}(t_1) \rangle.~~~~~\label{Gma-N}
\end{eqnarray}
which is related to an N-photon coincidence measurement. However,
this function does not provide any information about photon
entanglement, i.e., quantum superposition of different states.

The realization of multi-particle entanglement is paramount in
achieving most of the tasks in quantum computing and quantum
information processing \cite{sho,gro}. While there are many ways
to create entangled multi-particle state, the straightforward
method is to start from independent single photons \cite{hof}.
Knill, Laflamme, and Milburn \cite{klm} have shown that quantum
computing can be realized with single photons and some linear
optical elements via multi-photon interference. N-photon
entanglement is thus produced from single-photon states. This is
one of the primary reasons behind the big rush in creating light
sources with single-photon on demand \cite{mic2,san}.

While most of the analysis are based on the single mode model,
i.e., all the photons in one single temporal mode, this is, on the
other hand, impossible to achieve in experiment. The multi-mode
nature of light inevitably reduces the effect of photon
interference and leads to degradation in information processing.
One often uses the fidelity quantity of quantum states to
characterize the degradation. But this description has emphasis
only on the end result of the process and spares the true culprit
of the process, that is, the multi-mode nature of light.

For monochromatic field of only one frequency component, the field
can be represented by an infinite wave train. Photons can appear
anywhere in this wave train and are indistinguishable from each
other. They will produce maximum effect of entanglement. However,
when many frequency components are excited, an optical field is no
longer monochromatic and the wave train becomes finite with a
length of the order of the coherence length of the field. With
multiple photons, we generally cannot use a single wave packet to
describe them. We cannot assign separate wave packets to describe
each photon, either. This is because of the possibility of
multi-photon entanglement. Thus, an issue is raised about how to
describe the different situations of temporal distribution of
photons and distinguish these situations experimentally.

Recently, this issue was addressed in the four-photon case
\cite{tsu,ou2} for distinguishing a genuine four-photon
polarization entangled state from a state made of two well
separated pairs of photons. The difference lies in the
multi-photon interference: an entangled four-photon state will
give rise to the strongest multi-photon interference effect
whereas two well separate pairs produce less interference effect.
This is in consistence with the complementary principle of quantum
mechanics which states that quantum interference is a result of
indistinguishability of the paths but if the the paths are
distinguishable, the interference effect will be gone. Partial
distinguishability will lead to reduced interference effect, as
described by Eq.(\ref{2}) in the coherence theory for the
single-photon interference. Four-photon interference experiments
were performed to distinguish an entangled four-photon state from
two independent pairs of photons \cite{tsu,ou2,rhe1,rhe2}

However, the above mentioned interference scheme on the
four-photon state cannot be generalized to arbitrary photon
number. More recently, Sun {\it et al} \cite{sun1,sun2} and Resch
{\it et al} \cite{res} independently constructed a quantum state
projection measurement scheme and applied it to a maximally
entangled N-photon state (the so-called NOON state) for the
demonstration of multi-photon de Broglie wavelength without a NOON
state. It turns out that this new projection measurement scheme is
based on a multi-photon interference effect that depends on the
temporal distribution of the photons involved. Since the new
scheme can be easily generalized to arbitrary photon number, it
can be used to study the relation between the multi-photon
interference effect and the temporal distinguishability of an
N-photon state. We will show that the various scenarios of
temporal distribution of photons give rise to different visibility
in the multi-photon interference, which provides a direct measure
of the degree of temporal distinguishability of a multi-photon
state in a similar fashion to the coherence theory [Eq.(\ref{2})].
This is a quantitative investigation into the complementary
principle of quantum interference.

In the following, we will first review the two-photon and
four-photon cases to look for the relation between temporal
distinguishability and multi-photon interference. We then will
generalize to an arbitrary N-photon state and present the criteria
for photon indistinguishability and distinguishability. In
Sect.IV, we use quantum coherence theory to calculate the result
from a direct N-photon coincidence measurement and discuss the
generalized photon bunching effect. This measurement process is
not sensitive to the different temporal distribution of the
photons. In Sect.V, we introduce the newly constructed NOON state
projection measurement and demonstrate how it can be used to
characterize the degree of temporal indistinguishability for the
simple three-photon case. We will generalize the discussion for
three-photon case to the general N+1-photon case. In Sect.VI, we
will discuss an even more general case and present the numerical
results for a few special cases. We conclude with a discussion.

\section{Temporal distinguishability for the case of two photons
and for the case of two-pairs of photons}

The first discussion about the temporal distinguishability was by
Grice and Walmsley \cite{wam}, who investigated the visibility in
a Hong-Ou-Mandel interferometer \cite{hom} with a two-photon state
input from type-II parametric down-conversion. Later on, Atat\"ure
et al \cite{ser} performed experiment and confirmed the
degradation of the two-photon interference visibility predicted in
Ref.\cite{wam} due to temporal distinguishability.

In the discussion of Ref.\cite{wam}, the multi-mode description of
the two-photon state is given by
\begin{eqnarray}
|\Phi_{2}\rangle = \int
d\omega_1d\omega_2\Phi(\omega_1,\omega_2)\hat
a_s^{\dag}(\omega_1)\hat
a_i^{\dag}(\omega_2)|0\rangle,\label{phi-st}
\end{eqnarray}
where $s, i$ denote the two correlated signal and idler photons
from parametric down-conversion. For type-II process, we have
$\Phi(\omega_1,\omega_2) \ne \Phi(\omega_2,\omega_1)$ due the
birefringent effect of the nonlinear crystal on the ordinary and
extra-ordinary rays. The maximum visibility in the two-photon
Hong-Ou-Mandel interferometer has the form of
\begin{eqnarray}
{\cal V}_2 = M_2 \equiv {\int d\omega_1 d\omega_2 \Phi^*(\omega_1,
\omega_2)\Phi(\omega_2, \omega_1)\over \int d\omega_1 d\omega_2
|\Phi(\omega_1, \omega_2)|^2}.\label{9-2}
\end{eqnarray}
$M_2$ is defined as a degree of permutation symmetry. Note that
$M_2 =M_2^*$ and $0\le |M_2|\le 1$. The visibility or the degree
of permutation symmetry is one if and only if
$\Phi(\omega_1,\omega_2)$ satisfies the permutation symmetry
relation:
\begin{eqnarray}
\Phi(\omega_1, \omega_2)=\Phi(\omega_2, \omega_1).\label{9-3}
\end{eqnarray}
As stated in Ref.\cite{wam}, this permutation relation is a
signature of spectral indistinguishability of the two photons,
that is, we cannot tell the difference between the two photons
through their spectra. This in turn gives temporal
indistinguishability if we consider the Fourier transformation:
\begin{eqnarray}
G(t_1, t_2) = {1\over 2\pi}\int d\omega_1d\omega_2 \Phi(\omega_1,
\omega_2) e^{-i(\omega_1t_1+\omega_2t_2)}.\label{9-4}
\end{eqnarray}
Combination of Eqs.(\ref{9-3}, \ref{9-4}) gives directly the
symmetric relation:
\begin{eqnarray}
G(t_1, t_2)=G(t_2, t_1),\label{9-5}
\end{eqnarray}
for all times of $t_1, t_2$.

On the other hand, the visibility is zero if $\Phi(\omega_1,
\omega_2)$ does not have any overlap with $\Phi(\omega_2,
\omega_1)$, which is characterized by the orthogonal relation:
\begin{eqnarray}
\int d\omega_1 d\omega_2 \Phi^*(\omega_1, \omega_2)\Phi(\omega_2,
\omega_1) = 0\label{9-6}
\end{eqnarray}
or in time
\begin{eqnarray}
\int dt_1dt_2 G^*(t_1, t_2)G(t_2, t_1) = 0.\label{9-7}
\end{eqnarray}
This orthogonal relation indicates that the two functions $G(t_1,
t_2), G(t_2, t_1)$ have no overlap.

At this point, it is not easy to see what is the physical meaning
of Eq.(\ref{9-7}). However, if we go back to Eq.(\ref{9-3}) and
introduce a non-symmetric factor of $e^{i\omega_2T}$, we find that
the equivalent $\Phi(\omega_1, \omega_2)$ in Eq.(\ref{phi-st}) in
this case will be $\Phi^{\prime}(\omega_1, \omega_2)\equiv
\Phi(\omega_1, \omega_2)e^{i\omega_2T}$, which is not symmetric
with respect to $\omega_1,\omega_2$ even if $\Phi(\omega_1,
\omega_2)$ is symmetric. This extra phase can be introduced by
acting the evolution operator $\hat U(T) = \exp(-i\omega_2\hat
a_i^{\dag}\hat a_i T)$ on the state in Eq.(\ref{phi-st}) for an
extra free propagation time $T$ of the idler photon. This then
creates a time delay $T$ between the two photons. Then the
visibility in Eq.(\ref{9-2}) becomes
\begin{eqnarray}
{\cal V}_2(T) = {\int d\omega_1 d\omega_2 \Phi^*(\omega_1,
\omega_2)\Phi(\omega_2, \omega_1)e^{i(\omega_1-\omega_2)T}\over
\int d\omega_1 d\omega_2 |\Phi(\omega_1, \omega_2)|^2}.\label{v-p}
\end{eqnarray}
Notice that if the delay is large enough [$T >> T_c\sim
1/\Delta\omega_{PDC}$ with $\Delta\omega_{PDC}$ as the range of
$\Phi(\omega_1, \omega_2)$], we will have ${\cal V}_2(T) = 0$ or
$\Phi^{\prime}(\omega_1, \omega_2)$ satisfies Eq.(\ref{9-6}).
Since $T$ is the relative delay between the two photons before
they meet at the beam splitter of the Hong-Ou-Mandel
interferometer, we may believe that there is a large enough delay
between the two photons so that the two photons become
distinguishable in time when they arrive at the beam splitter.  So
the orthogonal relation in Eq.(\ref{9-6}) or Eq.(\ref{9-7})
corresponds to the situation when the two photons are well
separated in time and form two non-overlapping and distinguishable
wave packets.

Therefore, the visibility in the Hong-Ou-Mandel interferometer in
Eq.(\ref{9-2}) is a direct measure of temporal distinguishability
of the two photons. This is very much similar to the role of the
field correlation function $\gamma$ of Eq.(\ref{2}) in defining
optical coherence of a field.

For the four-photon case, temporal distinguishability between two
pairs of photons was first studied by Ou, Rhee and Wang
\cite{rhe1,rhe2} in a similar scheme as the Hong-Ou-Mandel
interferometer but with four photons. It was found that the
visibility in four-photon interference is directly related to a
quantity ${\cal E/A}$, which is a measure of the temporal
distinguishability of photon pairs from parametric
down-conversion: when ${\cal E/A} << 1$, the pairs are well
separated from each other corresponding to the so-called $2\times
2$ case but when ${\cal E/A}=1$, the two pairs are overlap in time
and form an indistinguishable four-photon state corresponding to
the $4\times 1$ case.

From the definition of the quantities ${\cal E}$ and ${\cal A}$ in
Ref.\cite{rhe2}, we rewrite them as
\begin{eqnarray}
&&{\cal E} = \int d\omega_1d\omega_2d\omega_1^{\prime}d
\omega_2^{\prime}\Phi^*(\omega_1, \omega_2)\Phi^*
(\omega_1^{\prime}, \omega_2^{\prime})\cr &&\hskip 1.2in \times
\Phi(\omega_1^{\prime}, \omega_2)\Phi(\omega_1,
\omega_2^{\prime}), \label{8-181}
\end{eqnarray}
and
\begin{eqnarray}
{\cal A} = \int d\omega_1d\omega_2d\omega_1^{\prime}d
\omega_2^{\prime}|\Phi(\omega_1, \omega_2)\Phi (\omega_1^{\prime},
\omega_2^{\prime})|^2, \label{8-182}
\end{eqnarray}
where $\Phi(\omega_1, \omega_2)$ is the two-photon wave function
in Eq.(\ref{phi-st}).

On the other hand, the four-photon state from Ref.\cite{rhe2} has
the form of
\begin{eqnarray}
|\Phi_{4}\rangle &=& \int d\omega_1d\omega_2d\omega_1^{\prime}d
\omega_2^{\prime}\Phi_4(\omega_1,\omega_2;\omega_1^{\prime},
\omega_2^{\prime})\cr &&\hskip 0.4in \times\hat
a_s^{\dag}(\omega_1)\hat a_i^{\dag}(\omega_2)\hat
a_s^{\dag}(\omega_1^{\prime})\hat
a_i^{\dag}(\omega_2^{\prime})|0\rangle,\label{phi4-st}
\end{eqnarray}
where $\Phi_4(\omega_1,\omega_2;\omega_1^{\prime},
\omega_2^{\prime})\equiv
\Phi(\omega_1,\omega_2)\Phi(\omega_1^{\prime},
\omega_2^{\prime})$. Then we can rewrite the expression for ${\cal
E}$ and ${\cal A}$ in Eqs.(\ref{8-181}, \ref{8-182}) and obtain
the quantity ${\cal E/A}$ as
\begin{widetext}
\begin{eqnarray}
{\cal E \over A} = {\int d\omega_1d\omega_2d\omega_1^{\prime}d
\omega_2^{\prime}\Phi_4^*(\omega_1, \omega_2,\omega_1^{\prime},
\omega_2^{\prime}) \Phi_4(\omega_1^{\prime}, \omega_2;\omega_1,
\omega_2^{\prime})\over \int d\omega_1d\omega_2d\omega_1^{\prime}d
\omega_2^{\prime}|\Phi_4(\omega_1, \omega_2;\omega_1^{\prime},
\omega_2^{\prime})|^2}.\label{e-a}
\end{eqnarray}
\end{widetext}
Recall that this quantity is a measure of the temporal
distinguishability of two pairs of photons. But from
Eq.(\ref{e-a}), we find that this quantity is again dependent on
the permutation of the wave function
$\Phi_4(\omega_1,\omega_2;\omega_1^{\prime}, \omega_2^{\prime})$
similar to that in Eq.(\ref{9-2}) and it is one if and only if we
have the permutation symmetry of
\begin{eqnarray}
\Phi_4(\omega_1, \omega_2,\omega_1^{\prime}, \omega_2^{\prime})
=\Phi_4(\omega_1^{\prime}, \omega_2;\omega_1,
\omega_2^{\prime}).\label{sym4}
\end{eqnarray}
Therefore from the discussion on the meaning of the quantity
${\cal E/A}$, we find that the symmetry relation in
Eq.(\ref{sym4}) corresponds to the case when the two pairs are
completely overlap in time and become temporally indistinguishable
(the $4\times 1$ case) whereas the orthogonal relation
\begin{eqnarray}
&&\int d\omega_1d\omega_2d\omega_1^{\prime}d
\omega_2^{\prime}\Phi_4^*(\omega_1, \omega_2,\omega_1^{\prime},
\omega_2^{\prime})\cr && \hskip 1in \times
\Phi_4(\omega_1^{\prime}, \omega_2;\omega_1,
\omega_2^{\prime})=0\label{orth4}
\end{eqnarray}
leads to the case of completely separated pairs of photons (the
$2\times 2$ case).

From the experiments and analysis on four-photon interference with
two pairs of photons by parametric down-conversion
\cite{rhe1,rhe2,sun1}, we find that the visibility is not zero
even for ${\cal E/A} =0$. This can be attributed to the existence
of two-photon interference since we usually have two-photon
indistinguishability with exchange symmetry in Eq.(\ref{9-3}).
Note that ${\cal E/A}$ concerns the permutation symmetry between
two different pairs, i.e., exchange between the group of
$\{\omega_1,\omega_2\}$ and the group of
$\{\omega_1^{\prime},\omega_2^{\prime}\}$. The exchange within
each group is symmetric due to Eq.(\ref{9-3}).

Next we will generalize Eqs.(\ref{9-3}, \ref{sym4}) and
Eqs.(\ref{9-6}, \ref{orth4}) to an arbitrary $N$-photon case and
relate them to the visibility of some $N$-photon interference
experiment.

\section{Description of a temporally distributed N-photon state}

Now we can generalize Eqs.(\ref{9-3}, \ref{9-6}) of the two-photon
case and Eqs.(\ref{sym4}, \ref{orth4}) of the two-pair case to
arbitrary $N$ case. An arbitrary N-photon state of wide spectral
range can be generally described by
\begin{eqnarray}
&&|\Phi_N\rangle = {\cal N}^{-1/2}\int
d\omega_1d\omega_2...d\omega_N \Phi(\omega_1, ...,
\omega_N)\times\cr &&\hskip 1.0in \times\hat
a^{\dag}(\omega_1)\hat a^{\dag}(\omega_2)...\hat
a^{\dag}(\omega_N)|0\rangle,\label{Phi-state}
\end{eqnarray}
where the normalization factor ${\cal N}$ is given by
\begin{eqnarray}
&&{\cal N} = \int d\omega_1d\omega_2...d\omega_N \Phi^*(\omega_1,
..., \omega_N)\times\cr &&\hskip 1.0in\times\sum_P
\Phi(P\{\omega_1, ..., \omega_N\}),\label{N}
\end{eqnarray}
where $P$ is the permutation operator on the indices of 1,2,...,
$N$. and the sum is over all possible permutation. There are
totally $N!$ terms. So the value of ${\cal N}$ ranges from $I$ to
$N!I$ with $I = \int d\omega_1d\omega_2...d\omega_N
|\Phi(\omega_1, ..., \omega_N)|^2$. The maximum value of $N!I$ is
reached when
\begin{eqnarray}
\Phi(\omega_1, ..., \omega_N) = \Phi(P\{\omega_1, ..., \omega_N\})
\end{eqnarray}
for all $P$. Similar to Eqs.(\ref{9-3}, \ref{sym4}), this
corresponds to a case when the $N$ photons are indistinguishable
in time. We refer to this case as the $N\times 1$ case, meaning
that all $N$ photons are in one indistinguishable temporal mode.
This single-mode description of an N-photon state is more vivid in
the special case when $\Phi(\omega_1, ..., \omega_N)$ is
factorized as $\phi(\omega_1)\phi(\omega_2)...\phi(\omega_N)$ and
the N-photon state simply becomes
\begin{eqnarray}
&&|\Phi_N\rangle = {1\over N!} \hat A(\phi)^{\dagger N}|0\rangle =
|N\rangle_{\phi}
\end{eqnarray}
with
\begin{eqnarray}
\hat A(\phi) = \int d\omega \phi(\omega)\hat a (\omega) ~~~~(\int
d\omega |\phi(\omega)|^2 = 1).
\end{eqnarray}
Note that $\hat A(\phi)$ satisfies $[\hat A, \hat A^{\dag}]=1$ and
represents the annihilation operator for a single temporal mode
characterized by $\phi(\omega)$. The single-photon state
$|1\rangle_{\phi}$ has a single-photon detection probability of
$|g(\tau)|^2$ with a temporal shape of
\begin{eqnarray}
g(\tau) = {1\over \sqrt{2\pi}}\int d\omega \phi(\omega)e^{-i\omega
t}\label{g}
\end{eqnarray}
and normalization relation
\begin{eqnarray}
\int d\tau|g(\tau) |^2 = 1.\label{g-norm}
\end{eqnarray}

The other extreme case of  ${\cal N} = I$ requires
$\Phi(\omega_1,...,\omega_N)$ be orthogonal to all the permuted
functions $\Phi(P\{\omega_1,...,\omega_N\})$ in the similar ways
in Eqs.(\ref{9-6}, \ref{orth4}) and thus corresponds to the
situation when all photons are well separated in time.  We refer
to this case as the $1\times N$ case, meaning that each photon is
in its separate temporal mode and there are totally $N$
independent modes.

For the situations in between the two extreme cases, the value of
${\cal N}$ is between $I$ and $N!I$. For example, assume that the
spectral amplitude $\Phi(\{\omega\})$ have partial permutation
symmetry, that is,
\begin{eqnarray}
\Phi(\omega_1, ..., \omega_N) = \Phi(P_{\{n_i\}}\{\omega_1, ...,
\omega_N\}),\label{sym}
\end{eqnarray}
where the permutation $P_{\{n_i\}}$ only applies to a subgroup of
$\{\omega_1, \omega_2,...,\omega_N\}$. In the meantime, it also
satisfies the orthogonal relations:
\begin{eqnarray}
\int d\omega_1...d\omega_N\Phi^*(\omega_1, ..., \omega_N)
\Phi(P_{ij}\{\omega_1, ..., \omega_N\})=0\label{orth}
\end{eqnarray}
for permutation $P_{ij}$ between different subgroups ($\{n_i\}$
and $\{n_j\}$, $i\ne j$) defined in Eq.(\ref{sym}). Then it can be
easily shown that ${\cal N} = n_1!n_2!...n_k!I$. In the simple
case when $\Phi(\omega)$ can be factorized as
\begin{eqnarray}
&&\Phi(\omega_1, ..., \omega_N)  =
\phi_1(\omega_1)...\phi_1(\omega_{n_1})
\phi_2(\omega_{n_1+1})\times \cr &&\hskip 1.2in
\times...\phi_2(\omega_{n_1+n_2}) ...\phi_k(\omega_N)~~~~~~
\end{eqnarray}
with the orthogonal relations
\begin{eqnarray}
\int d\omega_1d\omega_2\phi_i^*(\omega_1)\phi_j^*(\omega_2)
\phi_i(\omega_2)\phi_j(\omega_1)=0~~(i\ne j),\label{orth2}
\end{eqnarray}
the N-photon state in Eq.(\ref{Phi-state}) becomes
\begin{eqnarray}
&&|\Phi_N\rangle = {1\over n_1!}|n_1\rangle_{\phi_1} {1\over
n_2!}|n_2\rangle_{\phi_2}... {1\over
n_k!}|n_k\rangle_{\phi_k}.\label{Phi-state2}
\end{eqnarray}
This is the situation when the $N$ photons are divided into $k$
subgroups with $n_i (i=1,2,...,k)$ photons in each group in a
single temporal mode characterized by $\phi_i$. This situation is
denoted as $n_1+...+n_k$ case.

For simplicity of later argument, let us consider another special
kind of N-photon state with
\begin{eqnarray}
\Phi(\omega_1, ..., \omega_N) =
\phi(\omega_1)e^{i\omega_1T_1}...\phi(\omega_N)e^{i\omega_NT_N}.\label{Phi-T}
\end{eqnarray}
With this $\Phi$, the N-photon state can be viewed as direct
product of $N$ identical single photon wave packets:
\begin{eqnarray}
|N\rangle_T =
|T_1\rangle\otimes|T_2\rangle\otimes...\otimes|T_N\rangle,\label{Nprod}
\end{eqnarray}
with
\begin{eqnarray}
|T_i\rangle = \int d\omega \phi(\omega)e^{i\omega T_i}\hat
a^{\dag}(\omega)|0\rangle.
\end{eqnarray}
This state can be viewed as from single-photon sources such as
quantum dots (see below for details). However, the quantum state
in Eq.(\ref{Nprod}) is not normalized. Substituting
Eq.(\ref{Phi-T}) into Eq.(\ref{N}), we have the normalization
factor as
\begin{eqnarray}
&&{\cal N} = \int d\omega_1d\omega_2...d\omega_N
\Big[\prod_k|\phi(\omega_k)|^2e^{-i\omega_kT_k}\Big]\times\cr
&&\hskip 0.7in\times\sum_P P\Big[\exp\big\{\sum_m
i\omega_mT_m\big\}\Big].\label{NT}
\end{eqnarray}

When $T_1=T_2=...=T_N$, we recover the case when all $N$ photons
are in one single temporal mode with ${\cal N} =N!$ ($N\times 1$).
On the other hand, if $|T_i-T_j|>> 1/\Delta\omega (i\ne j)$ with
$\Delta\omega$ as the bandwidth of $\phi(\omega)$, we have ${\cal
N} =1$. This is the case when all the photons are well separated
from each other ($1\times N$).

The N-photon state in Eq.(\ref{Phi-state}) describes a state when
all photons are in one spatial and polarization mode. They only
differ in spectral mode. In practice, although $N=2$ case can be
easily obtained from degenerate parametric down-conversion, such a
state with $N >2$ is not easy to produce directly. It can be
produced indirectly from single-photon states with a set of beam
splitters as shown in Fig.1, where the single-photon sources are,
for example, quantum dots. The quantum state for the input fields
has the general form of
\begin{eqnarray}
&&|\Psi_N\rangle_{in} = \int d\omega_1d\omega_2...d\omega_N
\Phi(\omega_1, ..., \omega_N)\times\cr &&\hskip 1.0in \times\hat
a_1^{\dag}(\omega_1)\hat a_2^{\dag}(\omega_2)...\hat
a_N^{\dag}(\omega_N)|0\rangle ~~~~~~~\label{Phi-state-ND}
\end{eqnarray}
with the normalization relation:
\begin{eqnarray}
\int d\omega_1d\omega_2...d\omega_N \Phi^*(\omega_1, ...,
\omega_N)\Phi(\omega_1, ..., \omega_N) =1. \label{norm-ND}
\end{eqnarray}
Here $\hat a_j^{\dag} (j=1,...,N)$ is the creation operator for
each input mode. Photons are possible to exit at any of the $N$
output ports. To produce a state of the form in
Eq.(\ref{Phi-state}), however, we only consider the possibility
when all $N$ photon exit at one port, say, $b_N$ port. It is
straightforward using the beam splitter theory to show that the
projected state is
\begin{eqnarray}
&&\mathbb{P}|\Psi_N\rangle_{out} = {1\over N^{N/2}}\int
d\omega_1d\omega_2...d\omega_N \Phi(\omega_1, ...,
\omega_N)\times\cr &&\hskip 1.1in \times\hat
b_N^{\dag}(\omega_1)\hat b_N^{\dag}(\omega_2)...\hat
b_N^{\dag}(\omega_N)|0\rangle, ~~~~~~~\label{Phi-state-ND}
\end{eqnarray}
which is in the form of Eq.(\ref{Phi-state}). This state is not
normalized because it is a projected state with the probability of
projection as $P(|\Phi_N\rangle)
=||\mathbb{P}|\Psi_N\rangle_{out}||^2 = {\cal N}/N^N$. The delay
factors $\{e^{i\omega_jT_j}\}$ in Eq.(\ref{Phi-T}) can be easily
introduced on individual mode $\hat a_j$ before the beam splitters
via the free-field evolution operator $\hat U_j(T_j) =
\exp(-i\omega_j\hat a^{\dag}_j\hat a_j T_j)$.

\begin{figure}[htb]
\begin{center}
\includegraphics[width= 3in]{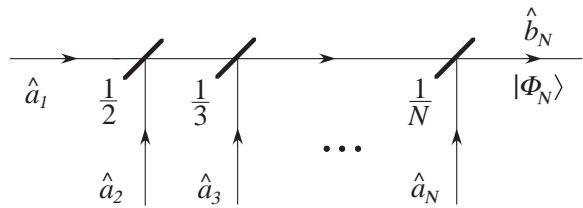}
\end{center}
\caption{\em Generation of an N-photon state from single-photon
states by beam splitters.} \label{fig1}
\end{figure}

More generally, to include different spatial and polarization
modes, the N-photon state has the following shape
\begin{widetext}
\begin{eqnarray}
|\Phi_N\rangle = {\cal N}_k^{-1/2}\int
d\omega_1^{(1)}...d\omega_{n_1}^{(1)}...d\omega_1^{(k)}...d\omega_{n_k}^{(k)}
\Phi(\{\omega^{(1)}\}, ..., \{\omega^{(k)}\})\hat
a^{\dag}_1(\omega_1^{(1)})...\hat
a^{\dag}_1(\omega_{n_1}^{(1)})...\hat
a^{\dag}_k(\omega_{n_k}^{(k)})|0\rangle,~~~~~~~~~\label{Phi-state-SP}
\end{eqnarray}
where $\{\omega^{(1)}\} =
\omega^{(1)}_1,...,\omega^{(1)}_{n_1}$, etc. The normalization
factor ${\cal N}_k$ takes the form of
\begin{eqnarray}
{\cal N}_k = \int d\{\omega^{(1)}\}...d\{\omega^{(k)}\}
\Phi^*(\{\omega^{(1)}\},...,\{\omega^{(k)}\})\sum_{P_1,...,P_k}
\Phi(P_1\{\omega^{(1)}\}, ..., P_k\{\omega^{(k)}\}).\label{Nk}
\end{eqnarray}
${\cal N}_k$ now ranges from $I$ to $n_1!...n_k!I$. The special
case when $\Phi(\{\omega^{(1)}\}, ..., \{\omega^{(k)}\})$
factorizes is similar as before.
\end{widetext}

\section{Direct N-Photon Measurement: Photon Bunching Effect for N Photons}

Next, we consider an N-photon joint measurement with the joint
probability density given from the quantum coherence theory in
Eq.(\ref{Gma-N}). The average is over the quantum state of the
system given in Eq.(\ref{Phi-state}) for an arbitrary N-photon
state. For simplicity, we first apply it to the state in
Eq.(\ref{Nprod}).

To carry out the quantum average, it is easier to first find the
N-photon detection probability amplitude:
\begin{eqnarray}
&&C^{(N)}(t_1,t_2,...,t_N) = \langle 0| \hat E^{(+)}(t_N)...\hat
E^{(+)}(t_1)|\Phi_N \rangle.~~~~~\label{CN}
\end{eqnarray}
Then $\Gamma^{(N)}(t_1,t_2,...,t_N) =
|C^{(N)}(t_1,t_2,...,t_N)|^2$. From Eq.(\ref{E}) for the field
operator and Eq.(\ref{Phi-T}) for $\Phi$, it is straightforward to
obtain
\begin{eqnarray}
&&C^{(N)}(t_1,t_2,...,t_N) \cr &&\hskip 0.5in = \sum_P
P[g(t_1-T_1)...g(t_N-T_N)],
\end{eqnarray}
where the permutation operation $P$ is on $t_1t_2...t_N$ and there
are $N!$ terms in the sum.

The overall probability of detecting $N$ photons together
(N-photon coincidence) is an integral of
$\Gamma^{(N)}(t_1,t_2,...,t_N)$ over all times $t_1,...,t_N$:
\begin{eqnarray}
&&P_N = \int dt_1...dt_N \Gamma^{(N)}(t_1,t_2,...,t_N) \cr
&&\hskip 0.23in = \int dt_1...dt_N \bigg|\sum_P
P[g(t_1-T_1)...g(t_N-T_N)]\bigg|^2.\cr &&\label{PN}
\end{eqnarray}

In the extreme case when $T_1 = T_2 = ... = T_N$, we obtain
$P_N{(N\times 1)} = (N!)^2 I$ while in the other extreme case when
$|T_i-T_j|>>1/\Delta\Omega$, we have $P_N{(1\times N)}= N!I$.
Therefore, we seem to have
\begin{eqnarray}
P_N{(N\times 1)} = N! P_N(1\times N),
\end{eqnarray}
or
\begin{eqnarray}
P_N{(N\times 1)}/ P_N(1\times N) = N!,
\end{eqnarray}
that is, the N-photon detection probability is $N!$ larger in the
case of $N$ identical photons than in the case of $N$ separated
photons. This can be thought of as the Bosonic photon bunching
effect for $N$ photons. The case of $N=2$ gives the familiar
photon bunching factor of 2.

However, as we know, the N-photon state in Eq.(\ref{Nprod}) is not
normalized. With the normalization factor considered, we have
instead
\begin{eqnarray}
P_N(N\times 1) = P_N(1\times N) = N!.\label{PN-N}
\end{eqnarray}
\begin{widetext}
For the case in between the two extreme cases, we may evaluate
Eq.(\ref{PN}) as

\begin{eqnarray}
P_N = \int dt_1...dt_N \sum_{P^{\prime}}
P^{\prime}[g^*(t_1-T_1)...g^*(t_N-T_N)] \sum_P
P[g(t_1-T_1)...g(t_N-T_N)]. \label{PN2}
\end{eqnarray}
Since the sum is over all permutations, the integral does not
change if we make the variable change: $\{t_1...t_N\}\rightarrow
P^{\prime}\{t_1...t_N\}$, i.e.,
\begin{eqnarray}
&&P_N = \sum_{P^{\prime}} \int dt_1...dt_N
g^*(t_1-T_1)...g^*(t_N-T_N) \sum_P P[g(t_1-T_1)...g(t_N-T_N)]\cr
&&\hskip 0.23in  = N! \int dt_1...dt_N g^*(t_1-T_1)...g^*(t_N-T_N)
\sum_P P[g(t_1-T_1)...g(t_N-T_N)]. \label{PN3}
\end{eqnarray}
It can be further shown that
\begin{eqnarray}
\int dt_1...dt_N g^*(t_1-T_1)...g^*(t_N-T_N) \sum_P
P[g(t_1-T_1)...g(t_N-T_N)]  = {\cal N},\label{PN4a}
\end{eqnarray}
\end{widetext}
where $\cal N$ is given in Eq.(\ref{NT}). Thus we
have
\begin{eqnarray}
P_N = N!\cal N.\label{PN4}
\end{eqnarray}
For a normalized N-photon state, we have $P_N = N!$ in any case as
in Eq.(\ref{PN-N}).

For the multi-spatial and polarization state in
Eq.(\ref{Phi-state-SP}), we may find $P_N$ after some lengthy
manipulation as that leads to Eq.(\ref{PN4}):
\begin{eqnarray}
P_N = n_1!...n_k!{\cal N}_k\label{PN5}
\end{eqnarray}
for the un-normalized state and $P_4= n_1!...n_k!$ for the
normalized state.

Hence, it is impossible to characterize different cases of
temporal entanglement with just simple direct multi-photon
detection for the normalized state. Furthermore, even for the
un-normalized state, we cannot explore the temporal
indistinguishability among different spatial and polarization
modes with multi-photon detection, for ${\cal N}_k$ depends only
on the permutation symmetry within photons in one spatial and
polarization mode.

Before we proceed further, it is interesting to evaluate the
multi-photon detection rates in some special cases. For example,
for the single-photon detection rate $P_1$, we have
\begin{eqnarray}
&&P_1 = \int dt \langle\Phi_N|\hat E^{\dagger}(t) \hat
E(t)|\Phi_N\rangle.
\end{eqnarray}
With some manipulation, it can be shown that $P_1 = N{\cal N}$ for
an un-normalized N-photon state and $P_1 = N$ for a normalized
N-photon state.

The reason that we still discuss the un-normalized case of an
N-photon state is because we encounter this kind of state in
practice when a projection measurement is involved such as that in
Fig.1. Consider, for example, a multi-photon state from degenerate
parametric down-conversion, which, for small $\eta$,  has the form
of \cite{sun1}
\begin{eqnarray}
&&|\Phi_{PDC}\rangle = C\Big(|0\rangle + \eta |\Phi_{2D}\rangle
+{\eta^2\over 2}|\Phi_{4D}\rangle +...\Big),\label{PDC}
\end{eqnarray}
with
\begin{eqnarray}
|\Phi_{2D}\rangle = \int
d\omega_1d\omega_2\Phi(\omega_1,\omega_2)\hat
a^{\dag}(\omega_1)\hat a^{\dag}(\omega_2)|0\rangle
\end{eqnarray}
and
\begin{eqnarray}
&&|\Phi_{4D}\rangle = \int d\omega_1d\omega_2d\omega_1'd\omega_2'
\Phi(\omega_1,\omega_2)\Phi(\omega_1',\omega_2')\times\cr&&\hskip
0.8in\times \hat a^{\dag}(\omega_1)\hat a^{\dag}(\omega_2)\hat
a^{\dag}(\omega_1')\hat a^{\dag}(\omega_2')|0\rangle.~~~~~~
\end{eqnarray}
Here $C$ in Eq.(\ref{PDC}) is a normalization factor but because
$|\eta|<< 1$, $|C|\approx 1$ no matter what function
$\Phi(\omega_1,\omega_2)$ is. Two-photon and four-photon
detections project the state to $\eta|\Phi_{2D}\rangle$ and
$\eta^2|\Phi_{4D}\rangle/2$, respectively, which are not
normalized. From Eqs.(\ref{N},\ref{PN4}), we then have
\begin{eqnarray}
&&P_2 = 2|\eta|^2\int
d\omega_1d\omega_2[|\Phi(\omega_1,\omega_2)|^2 +\cr&&\hskip
1in+\Phi^*(\omega_1,\omega_2)\Phi(\omega_2,\omega_1)].~~~~\label{P2}
\end{eqnarray}
The last term is related to the permutation symmetry or the degree
of two-photon temporal distinguishability and can be viewed as a
two-photon bunching effect. For a state from parametric
down-conversion in the degenerate case as in Eq.(\ref{PDC}), we
usually have the symmetry $\Phi(\omega_1,\omega_2) =
\Phi(\omega_2,\omega_1)$ so that
\begin{eqnarray}
&&P_{2D} = 4|\eta|^2\int
d\omega_1d\omega_2|\Phi(\omega_1,\omega_2)|^2 .~~~~\label{P2D}
\end{eqnarray}
Similarly for four-photon case, we have
\begin{eqnarray}
&&P_{4D} = 48|\eta|^2({\cal A}+2{\cal E}) = 3 P_2^2 (1+2{\cal
E/A}),~~~~\label{P4D}
\end{eqnarray}
where ${\cal E}, {\cal A}$ are given in Eqs.(\ref{8-181},
\ref{8-182}), respectively. The dependence on ${\cal E/A}$
indicates that the extra term in Eq.(\ref{P4D}) is a pair bunching
effect -- a generalized photon bunching effect for a multi-photon
state. Direct measurement by Sun {\it et al} \cite{sun2} confirmed
the four-photon bunching effect in Eq.(\ref{P4D}).

Another example is from non-degenerate parametric down-conversion
in type-II $\chi^{(2)}$ medium. The quantum state is similar to
that in Eq.(\ref{PDC}) \cite{rhe2}:
\begin{eqnarray}
&&|\Phi_{NPDC}\rangle = C\Big(|0\rangle + \eta |\Phi_{2N}\rangle
+{\eta^2\over 2}|\Phi_{4N}\rangle +...\Big),~~~~~\label{NPDC}
\end{eqnarray}
with
\begin{eqnarray}
|\Phi_{2N}\rangle = \int
d\omega_1d\omega_2\Phi(\omega_1,\omega_2)\hat
a^{\dag}_H(\omega_1)\hat a^{\dag}_V(\omega_2)|0\rangle
\end{eqnarray}
and
\begin{eqnarray}
&&|\Phi_{4N}\rangle = \int d\omega_1d\omega_2d\omega_1'd\omega_2'
\Phi(\omega_1,\omega_2)\Phi(\omega_1',\omega_2')\times\cr&&\hskip
0.8in\times \hat a^{\dag}_H(\omega_1)\hat a^{\dag}_V(\omega_2)\hat
a^{\dag}_H(\omega_1')\hat a^{\dag}_V(\omega_2')|0\rangle.~~~~~~
\end{eqnarray}
From Eq.(\ref{PN5}), it is straightforward to have
\begin{eqnarray}
&&P_{2N} = |\eta|^2\int
d\omega_1d\omega_2|\Phi(\omega_1,\omega_2)|^2 ~~~~\label{P2N'}
\end{eqnarray}
and
\begin{eqnarray}
&&P_{4N} = 2|\eta|^2({\cal A}+{\cal E}) = 2P_2^2 (1+{\cal
E/A}).~~~~\label{P4N}
\end{eqnarray}
Although there is no photon bunching at two-photon detection, we
still have the pair bunching effect that depends on the ${\cal
E/A}$ quantity.

\section{N-Photon Interference from an N-photon state}

As seen in the previous section, a direct N-photon detection
scheme cannot characterize the temporal indistinguishability in
the general case. Therefore, we need to seek another method. Since
the direct result of photon indistinguishability is the
interference effect, our scheme will be an N-photon interference
scheme. As a matter of fact, a Hong-Ou-Mandel interferometer
\cite{hom} has already been used to measure two-photon
indistinguishability from a type-II non-degenerate parametric
down-conversion \cite{wam,ser}. Our method proposed in the
following will be a generalization of the Hong-Ou-Mandel
interferometer from a two-photon case to an arbitrary N-photon
case.

\subsection{NOON State Projection as a Measure for
Distinguishability}

The NOON state projection measurement was recently proposed to
demonstrate an N-photon de Broglie wavelength without the need for
a NOON state \cite{sun1}. It was demonstrated for $N= 4$ with
states from parametric down-conversion \cite{sun2} and $N=6$ for a
coherent state \cite{res} experimentally . The scheme is depicted
in Fig.2 where the input is an arbitrary N-photon state of two
polarization modes in the form of
\begin{eqnarray}
&&|\Psi_N\rangle = \sum_{k=0}^N c_k |N-k,
k\rangle.~~~~\label{PsiN}
\end{eqnarray}
The N-photon coincidence probability from the $N$ detectors is
proportional to
\begin{eqnarray}
&&P_N \propto |\langle NOON|\Psi_N\rangle|^2.~~~~\label{P-N}
\end{eqnarray}
If the input state is of the form of $|N-k,k\rangle (k\ne0, N)$,
the output of the projection is zero. From the construction of the
NOON state, we find this orthogonal projection is a result of
N-photon interference and thus it can be used to characterize the
temporal indistinguishability by the visibility in the
interference. We will demonstrate this in the following sections.

\begin{figure}[htb]
\begin{center}
\includegraphics[width= 3in]{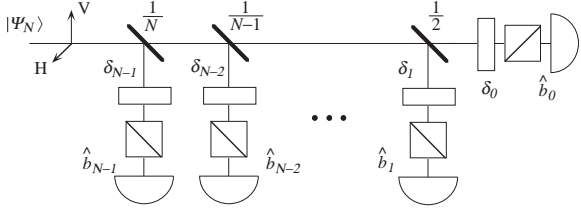}
\end{center}
\caption{\em A NOON-state projection measurement.
$\delta_k=2k\pi/N$ is the phase difference between H and V. $\hat
b_k\propto\hat E_H-\hat E_Ve^{i\delta_k}$.} \label{fig1}
\end{figure}

\subsection{Three-photon case}

Let us start with a three-photon state of the form $|2_H,
1_V\rangle$. So the three-photon NOON state projection measurement
should yield null three-photon coincidence in the ideal case when
all three photons are in one temporal mode. However, there may be
some delay between the vertical photon and the two horizontal
photons due to birefringence. Furthermore, the two horizontal
photons may also be separated from each other. To account for the
three scenarios described above, we cannot use the single-mode
state of $|2_H, 1_V\rangle$ and have to resort to the multi-mode
model discussed in Sect.III.

A multi-mode three-photon polarization state for $|2_H,
1_V\rangle$ has the form of
\begin{eqnarray}
&&|\Phi_{3}\rangle = \int d\omega_1d\omega_2d\omega_3
\Phi(\omega_1,\omega_2,\omega_3)\times\cr&&\hskip 0.8in\times \hat
a^{\dag}_H(\omega_1)\hat a^{\dag}_H(\omega_2)\hat
a^{\dag}_V(\omega_3)|0\rangle.~~~~~~\label{Phi3}
\end{eqnarray}
For simplicity of argument, we take
$\Phi(\omega_1,\omega_2,\omega_3)$ in the form of
Eq.(\ref{Phi-T}):
\begin{eqnarray}
&& \Phi(\omega_1,\omega_2,\omega_3)\cr&&\hskip
0.3in=\phi(\omega_1)e^{i\omega_1T_1}
\phi(\omega_2)e^{i\omega_2T_2}\phi(\omega_3)e^{i\omega_3T_3}.~~~~~~\label{Phi3-T}
\end{eqnarray}
We will use the un-normalized state because in practice, the state
in Eq.(\ref{Phi3}) can be generated by superposing a weak coherent
state $|\alpha\rangle$ with a two-photon state $|\eta\rangle =
|0\rangle+\eta|1_H,1_V\rangle$ from non-degenerate parametric
down-conversion:
\begin{eqnarray}
&&|\Phi_{3}\rangle = |\alpha\rangle_H|\eta\rangle \approx
|0\rangle + \alpha|1_H,0_V\rangle + \cr&&\hskip 0.5in
+(\alpha^2/\sqrt{2})|2_H,0_V\rangle +\eta|1_H,1_V\rangle +
\cr&&\hskip 0.6in + (\alpha^3/\sqrt{6})|3_H,0_V\rangle
+\eta\alpha|2_H,1_V\rangle,~~~~~~\label{Phi3SM}
\end{eqnarray}
where the states are in a single temporal mode and we only write
out states up to three photons. A three-photon coincidence measure
like that in the N-photon NOON state projection will only have
contributions from the last two terms. By making the coherent
state weak enough so that $|\eta|>>|\alpha|^2$, we are left with
only $|2_H,1_V\rangle$ term. Since $|\alpha|, |\eta|<<1$, the
three-photon state is not normalized.

For the scenarios presented in the beginning, we can relate them
to different values of  $T_1,T_2,T_3$. So $T_1=T_2=T_3$ is  for
the case of three photons all in one single temporal mode. When
$|T_3-T_1| >> 1/\Delta\omega, |T_3-T_2| >> 1/\Delta\omega$, the
V-photon is far from the two H-photons. When $|T_2-T_1| >>
1/\Delta\omega$, the two H-photons are far apart.

For the projection measurement in Fig.2 with $N=3$, we have the
electric field operators at three detectors as
\begin{eqnarray}
\begin{cases}
\hat E_0(t) = [\hat E_H(t) - \hat E_V(t)]/\sqrt{6}, \cr\hat E_1(t)
= [\hat E_H(t) - e^{i2\pi/3}\hat E_V(t)]/\sqrt{6}, \cr\hat E_2(t)
= [\hat E_H(t) - e^{i4\pi/3}\hat E_V(t)]/\sqrt{6}.\label{E3}
\end{cases}
\end{eqnarray}
To find the three-photon coincidence probability, we first
calculate the time correlation function
\begin{eqnarray}
&&\Gamma^{(3)}(t_1,t_2,t_3) \cr&&\hskip 0.2in = \langle \hat
E_0^{\dag}(t_3)\hat E_1^{\dag}(t_2) \hat E_2^{\dag}(t_1) \hat
E_2(t_1)\hat E_1(t_2) \hat E_0(t_3)\rangle.~~~~~~\label{Ga}
\end{eqnarray}
It is easy to calculate $\hat E_2(t_1)\hat E_1(t_2) \hat
E_0(t_3)|\Phi_3\rangle$:
\begin{widetext}
\begin{eqnarray}
\hat E_2(t_1)\hat E_1(t_2) \hat E_0(t_3)|\Phi_3\rangle= {-1\over
6\sqrt{6}} \bigg(\hat E_H\hat E_V\hat E_He^{i2\pi/3}+\hat E_V\hat
E_H\hat E_H+\hat E_H\hat E_H\hat
E_Ve^{i4\pi/3}\bigg)|\Phi_3\rangle.\label{E3}
\end{eqnarray}
Here we dropped the terms that have no contribution. The order of
the operators is kept for the time variables $t_3t_2t_1$. With the
state in Eq.(\ref{Phi3}) and $\Phi_3$ in Eq.(\ref{Phi3-T}), it is
straightforward to find
\begin{eqnarray}
&&\hat E_2(t_1)\hat E_1(t_2) \hat E_0(t_3)|\Phi_3\rangle =
{-1\over 6\sqrt{6}}
\bigg\{\Big[G(t_1,t_2,t_3)+G(t_2,t_1,t_3)\Big]e^{i4\pi/3} +
\Big[G(t_1,t_3,t_2)+G(t_3,t_1,t_2)\Big] e^{i2\pi/3}+\cr &&\hskip
3in + \Big[G(t_2,t_3,t_1)+
G(t_3,t_2,t_1)\Big]\bigg\}|0\rangle,\label{E3G}
\end{eqnarray}
\end{widetext}
where
\begin{eqnarray}
&&G(t_1,t_2,t_3) = {1\over \sqrt{(2\pi)^3}}\int d\omega_1
d\omega_2 d\omega_3 \Phi(\omega_1,\omega_2,\omega_3)\times\cr
&&\hskip 1.7in \times
e^{-i(\omega_1t_1+\omega_2t_2+\omega_3t_3)}\cr &&\hskip 0.73in=
g(t_1-T_1)g(t_2-T_2)g(t_3-T_3)~~~~~~~~~~\label{G3}
\end{eqnarray}
with $g(\tau)$ given in Eq.(\ref{g}). The three-photon joint
detection probability is an integral of the correlation function
in Eq.(\ref{Ga}) over all time variables:
\begin{eqnarray}
P_3 = \int dt_1dt_2dt_3 \Gamma^{(3)}(t_1,t_2,t_3).~~~~~~\label{P3}
\end{eqnarray}

We are now ready to discuss the three scenarios presented in the
beginning of this section. The interference effect is best
measured by the visibility which is usually defined as the
relative depth of modulation as compared to the situation when the
interference effect is zero. In the three scenarios, we find the
situation when the V-photon is far apart from the two H-photons
corresponds to the case of no interference, which sets the
reference line for evaluating the visibility defined by
\begin{eqnarray}
{\cal V}_3 ={| P_3-P_3(T_3=\infty)|\over P_3(T_3=\infty)}.
\label{V3}
\end{eqnarray}
Experimentally, we can scan $T_3$ from $T_3=\infty$ until we
observe the dip in $P_3$ and use Eq.(\ref{V3}) to calculate the
visibility.

Depending on the separation between the two H-photons, we actually
only have two distinct cases: ($i$) the two H-photons are
completely indistinguishable with $T_1=T_2\equiv T$; ($ii$) the
two H-photons are well separated and distinguishable in time with
$|T_1-T_2|>>1/\Delta\omega$.

In case ($i$) with $T_1=T_2\equiv T$, we have the exchange
symmetry $G(t_1,t_2,t_3) = G(t_2,t_1,t_3)$ and Eq.(\ref{P3})
becomes after the time integral
\begin{eqnarray}
P_3 =  2{{\cal A}_3-{\cal E}_3(\Delta T)\over 36}~~~~~~\label{P33}
\end{eqnarray}
with $\Delta T = T_3-T$ and
\begin{eqnarray}
{\cal A}_3 &\equiv & \int d\omega_1 d\omega_2 d\omega_3
|\Phi(\omega_1,\omega_2,\omega_3)|^2 \cr&=& \bigg(\int
d\omega|\phi(\omega)|^2\bigg)^3,~~~~~
\end{eqnarray}
\begin{eqnarray}
{\cal E}_3 (\tau) &\equiv & \int d\omega|\phi(\omega)|^2
\bigg|\int d\omega|\phi(\omega)|^2e^{-i\omega\tau}\bigg|^2.~~~~~
\end{eqnarray}
Note that ${\cal E}_3(0) ={\cal A}_3$ and ${\cal E}_3(\infty)=0$.
So from Eq.(\ref{V3}), we have the visibility for case ($i$) as
\begin{eqnarray}
{\cal V}_3(i) ={| P_3(\Delta T=0)-P_3(\Delta T=\infty)|\over
P_3(\Delta T=\infty)} = 1. \label{V3i}
\end{eqnarray}
The 100\% visibility corresponds to the single-mode discussion
before.

In case ($ii$) with $|T_1-T_2|>>1/\Delta\omega$, there is no
overlap between $G(t_1,t_2,t_3)$ and $G(t_2,t_1,t_3)$ so that
$\int dt_1dt_2dt_3 G^*(t_1,t_2,t_3)G(t_2,t_1,t_3) = 0$. We obtain
after the time integral
\begin{eqnarray}
P_3 =  {{\cal A}_3-{\cal E}_3(\Delta T_1)/2-{\cal E}_3(\Delta
T_2)/2\over 36}~~~~~~\label{P35}
\end{eqnarray}
with $\Delta T_1 = T_3 -T_1$ and $\Delta T_2 = T_3 -T_2$. So we
will have two dips with half depth when $T_3$ scans through $T_1$
and $T_2$. The visibility of each dip is then 50\%.

In summary, we find that the scenario when the two H-photons are
separated have a visibility of 50\% while when the two H-photons
are in one temporal mode, the interference visibility becomes
100\%. Therefore, we can distinguish the two different scenarios
in the three-photon case by measuring the visibility in the NOON
state projection measurement. Recent experiment by Liu {\it et al}
\cite{liu} realized the two scenarios described above and
confirmed the corresponding visibility. Next, we will generalize
this result to an N-photon state.

\subsection{$N+1$-photon case}

Let us now generalize the conclusion in the previous section to
the case of $|1_H,N_V\rangle$ with an arbitrary integer $N$. The
most general scenario in this case is when the single horizontal
photon (H) is indistinguishable from $m$ vertical photons (V)
while other $N-m$ V-photons are well separated in time from the
$m+1$ photons (the case of $1HmV+(N-m)V$ or $1HmV$ for short). The
multi-mode description of this state has the form of
\begin{widetext}
\begin{eqnarray}
|\Phi(1HmV)\rangle = \int d\omega_1 d\omega_2...d\omega_{N+1}
\Phi(\omega_1,..., \omega_N; \omega_{N+1})\hat
a_V^{\dag}(\omega_1)...\hat a_V^{\dag}(\omega_N)\hat
a_H^{\dag}(\omega_{N+1})|vac\rangle,~~~~~~\label{9-86}
\end{eqnarray}
with
\begin{eqnarray}
\Phi(\omega_1,..., \omega_N; \omega_{N+1})  =
\phi(\omega_1)e^{i\omega_1T_1} ... \phi(\omega_N)e^{i\omega_NT_N}
\phi(\omega_{N+1})e^{i\omega_{N+1}T_{N+1}}.~~~~~~~
\end{eqnarray}
Here we take $\Phi$ in the form of Eq.(\ref{Phi-T}) for ease of
calculation.
\end{widetext}

When $m$ H-photons are in the same temporal mode with the
V-photon, we have $T_1=...=T_m=T_{N+1}\equiv T$. But the other
$N-m$ V-photons are well separated from these $m+1$ photons. This
leads to $|T_j-T_k|>>1/\Delta\omega$ with $j=1,2,...,m, N+1$ and
$k=m+1,...,N$ and the orthogonal relation:
\begin{eqnarray}
&&\int dt_1dt_2 g^*(t_1-T_j)g^*(t_2-T_k)\cr && \hskip 0.6in \times
g(t_1-T_k)g(t_2-T_j) = 0.\label{9-90}
\end{eqnarray}

Now we are ready to evaluate the joint $N+1$-photon probability
$P_{N+1}$ in the NOON-state projection measurement scheme with an
input state of $|\Phi(1HmV)\rangle$  in Eq.(\ref{9-86}). $P_{N+1}$
is a time integral of the correlation function from $(N+1)$
detectors:
\begin{eqnarray}
&&\Gamma^{(N)}(t_1,t_2,...,t_N)  \cr &&\hskip 0.3in = \langle
\Phi(1HmV)|\hat E^{\dag}_{N+1}(t_{N+1})...\hat E^{\dag}_1(t_1) \cr
&& \hskip 0.8 in \times \hat E_1(t_1)...\hat E_{N+1}(t_{N+1})
|\Phi(1HmV)\rangle,~~~~~\label{9-91}
\end{eqnarray}
with
\begin{eqnarray}
\hat E_j(t) \propto \hat E_V(t) - \hat
E_H(t)e^{i\delta_j}+...,\label{9-92}
\end{eqnarray}
where
\begin{eqnarray}
\hat E_{H,V}(t) ={1\over \sqrt{2\pi}}\int d \omega \hat
a_{H,V}(\omega)e^{-i\omega t}.
\end{eqnarray}

It is easy to first evaluate $\hat E_1(t_1)...\hat
E_{N+1}(t_{N+1})$ $|\Phi(1HmV) \rangle$. After expanding the
product, we find only $N+1$ nonzero terms of the form
\begin{eqnarray}
-\sum_{k=1}^{N+1} e^{i \delta_k} \hat E_V(t_1)...\hat E_H(t_k)...
\hat E_V(t_{N+1})|\Phi(1HmV)\rangle.~~~~
\end{eqnarray}
For the state $|\Phi(1HmV)\rangle$ in Eq.(\ref{9-86}), we have
\begin{eqnarray}
&&\hat E_V(t_1)...\hat E_H(t_k)... \hat
E_V(t_{N+1})|\Phi(1HmV)\rangle \cr &&\hskip 0.8in ={\mathcal
G}(P_{k, N+1}\{t_1, ..., t_{N+1}\})|vac\rangle,~~~~~
\end{eqnarray}
with
\begin{eqnarray}
{\cal G}(t_1, ..., t_{N}; t_{N+1})= \sum_P G(P\{t_1, ..., t_{N}\};
t_{N+1}),\label{9-96}
\end{eqnarray}
and
\begin{eqnarray}
G(t_1, ..., t_{N}; t_{N+1})= \prod_{s=1}^{N+1}g(t_s-T_s),
\label{GN+1}
\end{eqnarray}
\begin{widetext}
\noindent where $P_{k, N+1}$ exchanges $t_k$ with $t_{N+1}$ and
$P$ is a permutation of $t_1, ..., t_{N}$. For the case of $1VmH$,
we have
\begin{eqnarray}
G(t_1, ..., t_{N+1})= g(t_{N+1}-T)
\prod_{s=1}^{m}g(t_s-T)\prod_{l=m+1}^{N}g(t_l-T_l), \label{G1VmH}
\end{eqnarray}
so that $G(t_1, ..., t_{N+1})$ has exchange symmetry in
$t_1,...,t_m, t_{N+1}$. The overall $(N+1)$-photon coincidence
probability is then given by
\begin{eqnarray}
P_{N+1}(1HmV) &\propto &\int dt_1...dt_{N+1} \bigg|
\sum_{k=1}^{N+1} e^{i\delta_k} {\cal G}(P_{k, N+1}\{t_1, ...,
t_{N+1}\})\bigg|^2\cr &=& \sum_{k,j} e^{i(\delta_k - \delta_j)}
\int dt_1...dt_{N+1} {\cal G}(P_{k, N+1}\{t_1, ..., t_{N+1}\})
{\cal G}^*(P_{j, N+1}\{t_1, ..., t_{N+1}\}).\label{9-97}
\end{eqnarray}
Diagonal terms of $k=j$ in the double sum are all
same because the integration is over all time variables:
\begin{eqnarray}
\int dt_1...dt_{N+1}   \big| {\cal G}(P_{k, N+1}\{t_1, ...,
t_{N+1}\})\big|^2 = \int dt_1...dt_{N+1} \big|{\cal G}(t_1, ...,
t_{N+1})\big|^2.
\end{eqnarray}
Furthermore,
\begin{eqnarray}
\int dt_1...dt_{N+1} \big|{\cal G}(t_1, ..., t_{N+1})\big|^2 &=&
\int dt_1...dt_Ndt_{N+1} \bigg|\sum_P
P[g(t_1-T_1)...g(t_N-T_N)]g(t_{N+1}-T_{N+1})\bigg|^2\cr &=& \int
dt_1...dt_N \bigg|\sum_P P[g(t_1-T_1)...g(t_N-T_N)]\bigg|^2,
\end{eqnarray}
where we used the normalization relation in Eq.(\ref{g-norm}).
From Eqs.(\ref{PN2}--\ref{PN4}), we find that it is simply
$N!{\cal N}$ with ${\cal N}$ given in Eq.(\ref{NT}). So the
diagonal terms of $k=j$ in Eq.(\ref{9-97}) are summed to be
$(N+1)N!{\cal N}$.

The cross terms in the double sum in Eq.(\ref{9-97}) are given by
\begin{eqnarray}
  \sum_{k\ne j}e^{i(\delta_k-\delta_j)} \int dt_1...dt_{N+1}  {\cal
G}(P_{k, N+1}\{t_1, ..., t_{N+1}\}) {\cal G}^*(P_{j, N+1}\{t_1,
..., t_{N+1}\}).
\end{eqnarray}
Let us consider one arbitrary term in the sum. The time integral
part can be rewritten as
\begin{eqnarray}
\sum_{P} \int dt_1...dt_{N+1} G(P\{t_1, ...,t_{k-1}, t_{N+1}
,t_{k+1},..., t_{N}\}; t_k) \sum_{P^{\prime}} G^*(P^{\prime}\{t_1,
...,t_{j-1}, t_{N+1} ,t_{j+1} , ..., t_{N}\}; t_j).
\end{eqnarray}
Since $k\ne j$, the variable set $\{t_1, ...,t_{k-1}, t_{N+1},
t_{k+1}$, $..., t_{N}\}$ is different from $\{t_1, ...,t_{j-1},
t_{N+1} ,t_{j+1} , ..., t_{N}\}$ only at $t_j$ and $t_k$. For
those $P$s such that $P\{t_1, ...,t_{k-1}, t_{N+1} ,t_{k+1},...,
t_{N}\}$ moves $t_{j}$ to the first $m$ positions in the variable
set $\{t_1, ..., t_{N}\}$, the symmetry between $t_1,...,t_m$ and
$t_{N+1}$ in the function $G(t_1, ..., t_{N}; t_{N+1})$ in
Eq.(\ref{G1VmH}) will make $G(P\{t_1, ...,t_{k-1}, t_{N+1}
,t_{k+1},..., t_{N}\}; t_k) = G(P\{t_1, ...,t_{j-1}, t_{N+1}
,t_{j+1} , ..., t_{N}\}; t_j)$. There are totally $m(N-1)!$ such
permutations and they all lead the time integral to
\begin{eqnarray}
\int dt_1...dt_{N+1} G(t_1, ...,t_{j-1}, t_{N+1} ,t_{j+1},...,
t_{N}\}; t_j) \sum_{P^{\prime}}G^*(P^{\prime}\{t_1, ...,t_{j-1},
t_{N+1} ,t_{j+1} , ..., t_{N}\}; t_j).
\end{eqnarray}
By Eq.(\ref{PN4a}), it is simply ${\cal N}$.

For the other permutations that move $t_j$ to the position of
$t_{m+1}, ..., t_N$, it cannot be interchanged with $t_k$ because
$T \ne T_s (s=m+1, ..., N)$.  Furthermore, by the orthogonal
relation in Eq.(\ref{9-90}), the time integral is simply zero.
Therefore, the cross terms are equal to
\begin{eqnarray}
\int dt_1...dt_{N+1} \sum_{k\ne j}e^{i(\delta_k-\delta_j)} {\cal
G}(P_{1k}\{t_1, ..., t_{N+1}\}) {\cal G}^*(P_{1j}\{t_1, ...,
t_{N+1}\}) = m (N-1)!{\cal N} \sum_{k\ne j}
e^{i(\delta_k-\delta_j)}.\label{9-102}
\end{eqnarray}
\end{widetext}
But because $\sum_k e^{i\delta_k} = 0$, we have
\begin{eqnarray}
\sum_{k\ne j} e^{i(\delta_k-\delta_j)} &&= \bigg(\sum_{k,j} -
\sum_{k=j}\bigg) e^{i(\delta_k-\delta_j)} \cr &&=  \sum_k
e^{i\delta_k} \sum_j e^{-i\delta_j} -(N+1) \cr
&&=-(N+1).\label{9-103}
\end{eqnarray}
So the final result is
\begin{eqnarray}
P_{N+1}(1HmV) &\propto & {\cal N} (N+1)(N-1)!(N- m ) \cr &=&
(N+1)!{\cal N}\Big(1-{m\over N}\Big).~~~~~\label{9-104}
\end{eqnarray}
For the generalized Hong-Ou-Mandel interferometer, we scan the
delay of the H-photon relative to the V-photons. When it does not
overlap with any of the V-photons, no interference occurs and
$P_{N+1}$ is a straight line which corresponds to $m=0$ in
Eq.(\ref{9-104}) with $P_{N+1}(\infty) = (N+1)!{\cal N}$. The
value in Eq.(\ref{9-104}) corresponds to the case when the delay
is zero between the $m$ V-photons and the one H-photon and a local
maximum interference is achieved. So the visibility is
\begin{eqnarray}
{\cal V}_{N+1}(1HmV) &\equiv &{P_{N+1}(\infty) - P_{N+1}(1HmV)
\over P_{N+1}(\infty)} \cr &=& {m\over N}.\label{VN+1}
\end{eqnarray}
Note that this visibility only depends on $N$ and $m$, i.e., the
total number $N$ of V-photons and the number $m$ of V-photons that
overlap with the single H-photon. It is independent of the
normalization factor ${\cal N}$ or how the other $N-m$ photons
distribute in time.

\begin{figure}
\includegraphics[width =
2.9in]{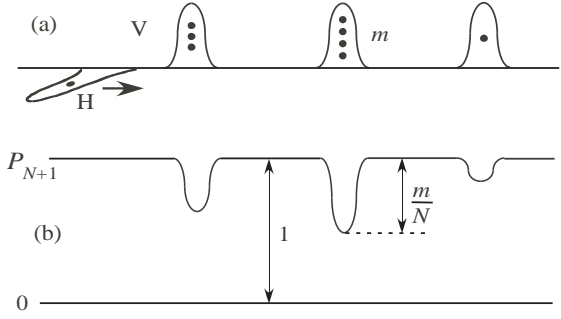} \caption[width=4in]{(a) A temporal distribution
with well separated groups of V-photons and (b) the corresponding
normalized $P_{N+1}$ as the position of the H-photon is scanned.}
\end{figure}

So for a temporal distribution of well separated groups of
V-photons shown in Fig.3a, as we scan the location of the single
H-photon, we will have more dips of various visibility (Fig.3b)
and the visibility is $m/N$ when the single H-photon overlaps with
the group of $m$ V-photons that are in one temporal mode and are
well separated from other V-photons.

In general for a temporal distribution with $m$ partially
overlapping V-photons, the visibility will be a value less than
$m/N$. Therefore, the experimentally measurable visibility of the
dips can be used to characterize the degree of temporal
indistinguishability of an N-photon state.

\section{The general case of $|k_H, N_V\rangle$ with $k>1$}

For a more general case of input state of $|k_H, N_V\rangle$ with
$k>1$, there are many scenarios for the temporal distribution of
the photons. We will start with the four-photon case of $k=N=2$.

\subsection{Four-photon case of $|2_V, 2_H\rangle$}

This situation was discussed in Ref.\cite{sun1} for $4\times 1$
case and $2HV\times 2HV$ case. It was shown that ${\cal V}_4
(4\times 1) =1$ and ${\cal V}_4 (2\times 2) =1/3$. But there are
other scenarios like $2H1V+1V$ and $1H1V+1H+1V$. We will consider
a simpler model to include these two scenarios so as to complete
the distinguishability discussion in the four-photon case.

For simplicity, we will again only discuss an un-normalized
independent four-photon state of the form
\begin{eqnarray}
&&|\Phi_{4}\rangle = \int d\omega_1d\omega_2d\omega_3d\omega_4
\Phi(\omega_1,\omega_2,\omega_3,\omega_4)\times\cr&&\hskip
0.8in\times \hat a^{\dag}_H(\omega_1)\hat a^{\dag}_H(\omega_2)\hat
a^{\dag}_V(\omega_3)a^{\dag}_V(\omega_4)|0\rangle.~~~~~~\label{Phi4}
\end{eqnarray}
with $\Phi(\omega_1,\omega_2,\omega_3,\omega_4)$ in the form of
Eq.(\ref{Phi-T}):
\begin{eqnarray}
&&
\Phi(\omega_1,\omega_2,\omega_3,\omega_4)=\phi(\omega_1)e^{i\omega_1T_1}
\phi(\omega_2)e^{i\omega_2T_2}\times\cr&&\hskip
1.2in\times\phi(\omega_3)e^{i\omega_3T_3}
\phi(\omega_4)e^{i\omega_4T_4}.~~~~~~\label{Phi4-T}
\end{eqnarray}

For the NOON state projection measurement with $N=4$, the field
operators at the four detectors are related to the input field
operators as
\begin{eqnarray}
\begin{cases}
\hat E_0(t) = [\hat E_H(t) - \hat E_V(t)]/2+... ,\cr\hat E_1(t) =
[\hat E_H(t) +\hat E_V(t)]/2+..., \cr\hat E_2(t) = [\hat E_H(t) -
i\hat E_V(t)]/2+..., \cr\hat E_3(t) = [\hat E_H(t) + i\hat
E_V(t)]/2+...,~~~~~~\label{E4}
\end{cases}
\end{eqnarray}
where we omit the vacuum modes. The four-photon detection
probability at the four detectors is related to the following
correlation function:
\begin{widetext}
\begin{eqnarray}
\Gamma^{(4)}(t_1,t_2,t_3,t_4) = \langle \hat E_0^{\dag}(t_4)\hat
E_1^{\dag}(t_3) \hat E_2^{\dag}(t_2) \hat E_3^{\dag}(t_1)\hat
E_3(t_1) \hat E_2(t_2)\hat E_1(t_3) \hat
E_0(t_4)\rangle.~~~~~~\label{Ga4}
\end{eqnarray}
Again, it is easy to first calculate $\hat E_3(t_1) \hat E_2(t_2)
\hat E_1(t_3)\hat E_0(t_4) |\Phi_4\rangle$. For this, we expand
$\hat E_3(t_1) \hat E_2(t_2) \hat E_1(t_3)\hat E_0(t_4)$:
\begin{eqnarray}
\hat E_3(t_1)\hat E_2(t_2)\hat E_1(t_3)\hat E_0(t_4) =
[(VVHH-HHVV) + i(VHVH+HVHV) -i(HVVH+VHHV)]/16, ~~~~~\label{EHV}
\end{eqnarray}
where $H=\hat E_H, V=\hat E_V$ and we keep the time ordering. For
the state $|\Phi_4\rangle$ in Eq.(\ref{Phi4}), we have
\begin{eqnarray}
HHVV|\Phi_4\rangle  = [G(t_1,t_2,t_3,t_4)+ G(t_2,t_1,t_3,t_4)+
G(t_1,t_2,t_4,t_3)+G(t_2,t_1,t_4,t_3)]|0\rangle,
~~~~~~\label{HHVV}
\end{eqnarray}
\begin{eqnarray}
VVHH|\Phi_4\rangle = [G(t_3,t_4,t_1,t_2)+ G(t_4,t_3,t_1,t_2)+
[G(t_3,t_4,t_2,t_1)+G(t_4,t_3,t_2,t_1)]|0\rangle,
~~~~~~\label{VVHH}
\end{eqnarray}
\begin{eqnarray}
HVHV|\Phi_4\rangle = [G(t_1,t_3,t_2,t_4)+ G(t_3,t_1,t_2,t_4)+
G(t_1,t_3,t_4,t_2)+G(t_3,t_1,t_4,t_2)]|0\rangle,
~~~~~~\label{HVHV}
\end{eqnarray}
\begin{eqnarray}
VHVH|\Phi_4\rangle  = [G(t_2,t_4,t_1,t_3)+ G(t_2,t_4,t_3,t_1)+
G(t_4,t_2,t_1,t_3)+G(t_4,t_2,t_3,t_1)]|0\rangle,
~~~~~~\label{VHVH}
\end{eqnarray}
\begin{eqnarray}
HVVH|\Phi_4\rangle = [G(t_1,t_4,t_2,t_3)+ G(t_1,t_4,t_3,t_2)+
G(t_4,t_1,t_2,t_3)+G(t_4,t_1,t_3,t_2)]|0\rangle,
~~~~~~\label{HVHV}
\end{eqnarray}
\begin{eqnarray}
VHHV|\Phi_4\rangle = [G(t_2,t_3,t_1,t_4)+ G(t_2,t_3,t_4,t_1)+
G(t_3,t_2,t_1,t_4)+G(t_3,t_2,t_4,t_1)]|0\rangle ~~~~~~\label{VHHV}
\end{eqnarray}
with
\begin{eqnarray}
G(t_1,t_2,t_3,t_4)= {1\over (2\pi)^2}\int d\omega_1
d\omega_2d\omega_3d\omega_4
\Phi(\omega_1,\omega_2,\omega_3,\omega_4)\times
e^{-i(\omega_1t_1+\omega_2t_2+\omega_3t_3+\omega_4t_4)}.~~~~~~\label{G4}
\end{eqnarray}
\end{widetext}
For the $\Phi$-function given in Eq.(\ref{Phi4-T}), the above
$G$-function is simply
\begin{eqnarray}
&&G(t_1,t_2,t_3,t_4)\cr &&\hskip 0.2 in=
g(t_1-T_1)g(t_2-T_2)g(t_3-T_3)g(t_4-T_4).~~~~~~\label{G4-T}
\end{eqnarray}
Four-photon coincidence probability is proportional to a time
integral of the correlation function $\Gamma^{(4)}$:
\begin{eqnarray}
P_4 = \int dt_1dt_2dt_3dt_4
\Gamma^{(4)}(t_1,t_2,t_3,t_4).~~~~~~\label{P4}
\end{eqnarray}

Next, we will evaluate $P_4$ for various scenarios of photon
distinguishability. To describe the four scenarios discussed in
the beginning of this section, we introduce three delay
parameters: $\Delta T, \Delta T_V, \Delta T_H$ so that $T_2 =
T_1+\Delta T_H, T_3 = T_1+\Delta T, T_4=T_3+\Delta T_V$.
Therefore, $\Delta T$ is for the delay between the H-photons and
the V-photons and $\Delta T_{H(V)}$ for the delay between the two
H(V)-photons. When $\Delta T=\pm\infty$, there is no overlap
between the H- and V-photons and no interference occurs. This sets
up the baseline for evaluating the visibility of interference. We
start with the $4\times 1$ case:

\noindent ({\em i}) $\Delta T_H =0 = \Delta T_V$. There is an
exchange symmetry between $t_1, t_2$ and between $t_3,t_4$ in
$G(t_1,t_2,t_3,t_4)$ with
\begin{widetext}
\begin{eqnarray}
G(t_1,t_2,t_3,t_4)= g(t_1-T_1)g(t_2-T_1)g(t_3-T_1-\Delta
T)g(t_4-T_1-\Delta T).~~~~~~\label{G4-Ti}
\end{eqnarray}
So we have
\begin{eqnarray}
&&\hat E_3(t_1)\hat E_2(t_2)\hat E_1(t_3)\hat
E_0(t_4)|\Phi_4\rangle ={1\over4}\Big
\{[G(t_1,t_2,t_3,t_4)-G(t_3,t_4, t_1,t_2)] +
i[G(t_1,t_3,t_2,t_4)+G(t_2,t_4, t_1,t_3)] \cr &&\hskip 2.8in
-i[G(t_1,t_4,t_3,t_2)+G(t_3,t_2, t_1,t_4)]\Big\}|0\rangle.
~~~~\label{E-4}
\end{eqnarray}
\end{widetext}
After the time integral, we obtain
\begin{eqnarray}
P_4(\Delta T) = {1\over 8}\big[3{\cal A}_4 -4{\cal
E}_4^{(1)}(\Delta T) + {\cal E}_4^{(2)}(\Delta T)\big]
\end{eqnarray}
with
\begin{eqnarray}
{\cal A}_4 = \bigg(\int d\omega|\phi(\omega)|^2\bigg)^4,
\end{eqnarray}
\begin{eqnarray}
{\cal E}_4^{(1)}(\tau) = \bigg(\int
d\omega|\phi(\omega)|^2e^{-i\omega\tau}\int
d\omega|\phi(\omega)|^2\bigg)^2,
\end{eqnarray}
\begin{eqnarray}
{\cal E}_4^{(2)}(\tau) = \bigg(\int
d\omega|\phi(\omega)|^2e^{-i\omega\tau}\bigg)^4.
\end{eqnarray}
Note that ${\cal E}_4^{(1)}(0)={\cal E}_4^{(2)}(0)={\cal A}_4$ and
${\cal E}_4^{(1)}(\infty)={\cal E}_4^{(2)}(\infty)=0$. As we scan
the relative delay $\Delta T$ between the H- and V-photons, the
four-photon coincidence will show an interference dip all the way
to zero when $\Delta T=0$, which corresponds to the case of
$T_1=T_2=T_3=T_4$ or the $4\times 1$ case. So the visibility is
100\% for the $4\times 1$ case.

\begin{widetext}
\noindent ({\em ii}) $\Delta T_H=0$ but $\Delta T_V
>>1/\Delta\omega$. In this case, the two V-photons are well separated and we have
\begin{eqnarray}
&&G(t_1,t_2,t_3,t_4)= g(t_1-T_1)g(t_2-T_1) g(t_3-T_1-\Delta T)
g(t_4-T_1-\Delta T_V-\Delta T).~~~~~~\label{G4-Tii}
\end{eqnarray}
When $\Delta T=0$, there is an exchange symmetry between
$\{t_1,t_2,t_3\}$ in $G(t_1,t_2,t_3,t_4)$. This is the $2H1V+1V$
case. But for arbitrary $\Delta T$, there is only a permutation
symmetry between $t_1,t_2$ in $G(t_1,t_2,t_3,t_4)$. Then we have
\begin{eqnarray}
&&\hat E_3(t_1)\hat E_2(t_2)\hat E_1(t_3)\hat
E_0(t_4)|\Phi_4\rangle \cr &&\hskip 0.5in={1\over8}\Big
\{[G(t_1,t_2,t_3,t_4)+ G(t_1,t_2,t_4,t_3)-G(t_3,t_4,
t_1,t_2)-G(t_3,t_4, t_2,t_1)] \cr &&\hskip 1in +
i[G(t_1,t_3,t_2,t_4)+G(t_1,t_3,t_4,t_2)+G(t_2,t_4,
t_1,t_3)+G(t_2,t_4, t_3,t_1)] \cr &&\hskip
1.2in-i[G(t_1,t_4,t_3,t_2)+G(t_3,t_2,
t_1,t_4,)+G(t_1,t_4,t_2,t_3)+G(t_3,t_2, t_4,t_1,)]\Big\}|0\rangle.
~~~~\label{E-4ii}
\end{eqnarray}
\end{widetext}
When $\Delta T=\pm\infty$, there is no overlap between all the
terms in Eq.(\ref{E-4ii}) so that all the cross terms are zero
after the time integral in Eq.(\ref{P4}).  So we have
\begin{eqnarray}
P_4(\Delta T=\pm\infty) = 3{\cal A}_4/16.
\end{eqnarray}
On the other hand, when $\Delta T=0$, there is an exchange
symmetry between $\{t_1,t_2,t_3\}$ in $G(t_1,t_2,t_3,t_4)$. So
Eq.(\ref{E-4ii}) becomes
\begin{eqnarray}
&&\hat E_3(t_1)\hat E_2(t_2)\hat E_1(t_3)\hat
E_0(t_4)|\Phi_4\rangle \cr &&\hskip 0.2in={1\over8}\Big
[G(t_1,t_2,t_3,t_4)+ G(t_1,t_2,t_4,t_3)\cr &&\hskip
0.5in-G(t_3,t_4, t_1,t_2)-G(t_3,t_4, t_2,t_1) \Big]|0\rangle
~~~~\label{E-4ii0}
\end{eqnarray}
and there is no overlap between all four terms above. After the
time integral, we obtain
\begin{eqnarray}
P_4(\Delta T=0) = {\cal A}_4/16.
\end{eqnarray}
So the visibility is
\begin{eqnarray}
{\cal V}_4(2H1V+1V) = 2/3.
\end{eqnarray}
for the $2H1V+1V$ case. In fact, there is another $2H1V+1V$ case
when the two H-photons overlaps with the other V-photon and
$\Delta T=-\Delta T_V$. In this case, we have the exchange
symmetry between $\{t_1,t_2,t_4\}$ in $G(t_1,t_2,t_3,t_4)$ so that
\begin{eqnarray}
P_4(\Delta T=-\Delta T_V) = {\cal A}_4/16,
\end{eqnarray}
which also gives ${\cal V}_4(2H1V+1V)=2/3$.

\noindent ({\em iii}) $\Delta T_H=\Delta T_V \equiv T
>>1/\Delta\omega$. This is the $1H1V+1H1V$ or the $2\times 2$ case when $\Delta T=0$
and we have the exchange symmetry between $\{t_1,t_3\}$ and
between $\{t_2,t_4\}$. But for $\Delta T=\pm\infty$, there is no
overlap between any two of the 24 terms in
Eqs.(\ref{HHVV}--\ref{VHHV}). So we have after the time integral
in Eq.(\ref{P4})
\begin{eqnarray}
P_4(\Delta T=\pm\infty) = 24{\cal A}_4/16^2 = 3{\cal
A}_4/32.\label{infty}
\end{eqnarray}
When $\Delta T=0$, on the other hand, we have
\begin{widetext}
\begin{eqnarray}
&&\hat E_3(t_1)\hat E_2(t_2)\hat E_1(t_3)\hat
E_0(t_4)|\Phi_4\rangle ={i\over 8}\Big
[G(t_1,t_3,t_4,t_2)+G(t_3,t_1, t_2,t_4)-G(t_1,t_4,
t_3,t_2)-G(t_4,t_1,t_2,t_3)\Big]|0\rangle. ~~~~~~~\label{E-4iii}
\end{eqnarray}
\end{widetext}
The above four terms have no overlap so that we obtain
\begin{eqnarray}
P_4(\Delta T=0) = 4{\cal A}_4/8^2 = {\cal A}_4/16.
\end{eqnarray}
Therefore, the visibility for the $2\times2$ case is simply
\begin{eqnarray}
{\cal V}_4(2\times2) = 1/3.
\end{eqnarray}

\noindent ({\em iv}) $|\Delta T_H-\Delta T_V|>>1/\Delta\omega$ and
$|\Delta T_H|, |\Delta T_V|>>1/\Delta\omega$. As we scan $\Delta
T$, there is an exchange symmetry only in one pair of the
variables between $\{t_1,t_2\}$ and $\{t_3,t_4\}$, that is,
between $\{t_1,t_3\}$ when $\Delta T=0$, or between $\{t_1,t_4\}$
when $\Delta T=-\Delta T_V$, or between $\{t_2,t_3\}$ when $\Delta
T=\Delta T_H$, or between $\{t_2,t_4\}$ when $\Delta T=\Delta
T_H-\Delta T_V$. This is the $(1H1V+1H+1V)$ case. In all these
cases, 8 out of 24 terms in Eqs.(\ref{HHVV}-\ref{VHHV}) are
cancelled in Eq.(\ref{EHV}) and the remaining ones are orthogonal
to each other so that we have
\begin{eqnarray}
P_4(\Delta T=0) = 16{\cal A}_4/16^2 = {\cal A}_4/16.
\end{eqnarray}
The situation when $\Delta T=\pm \infty$ is same as
Eq.(\ref{infty}). Therefore the visibility is
\begin{eqnarray}
{\cal V}_4(1H1V+1H+1V) = 1/3.
\end{eqnarray}
for the $1H1V+1H+1V$ case.

These are all likely distinct scenarios. We summarize the
visibility in Table I. Although visibility is derived with a
specific $\Phi$-function in Eq.(\ref{Phi4-T}), in general,
visibility is the same regardless of the form of $\Phi$ as long as
it is such that $G(t_1,t_2,t_3,t_4)$ has the required exchange
symmetry in each scenario listed above. The intermediate
situations will not have any symmetry in $G(t_1,t_2,t_3,t_4)$ and
thus have very complicated dependence on the various permutations
of $\Phi(\omega_1,\omega_2,\omega_3,\omega_4)$. Ref.\cite{sun1}
discussed the intermediate scenario from the $2\times 2$ case to
the $4\times 1$ case. Indeed, the visibility depends on the
quantity ${\cal E/A}$, which defines the degree of pair
distinguishability. Xiang {\it et al} \cite{xia} realized the
$2\times 2$ and the $4\times 1$ cases experimentally and confirmed
the visibility in Table I.

\begin{table}
\caption{\label{tab:table1}Visibility for 2 H-photons and 2
V-photons input }
\begin{ruledtabular}
\begin{tabular}{ccccccc}
\noalign{\smallskip} 2H2V&&2H1V+1V &&1HV+1HV&&1HV+H+V\\
\noalign{\smallskip} \hline \noalign{\smallskip} 1 && 2/3 &&
1/3&&1/3\smallskip
\end{tabular}
\end{ruledtabular}
\end{table}

\subsection{The Special Cases of $|2_H,
3_V\rangle$, $|2_H, 4_V\rangle$, and $|3_H, 3_V\rangle$}

Following the same line of derivation but in a much more
complicated fashion, we may find the visibility for all the
scenarios for the input states of $|2_H, 3_V\rangle$, $|2_H,
4_V\rangle$, and $|3_H, 3_V\rangle$. We list the likely scenarios
below and tabulate the visibility for each scenarios in Tables
II-IV.

\subsubsection{The Case of $|2_H, 3_V\rangle$}

The case of $|2_H, 3_V\rangle$ has 8 different scenarios. They are

\noindent (i) $~|2H3V\rangle, |2H2V+1V\rangle, |2H1V+2V\rangle$,
and

\noindent (ii) $|1H3V+1H\rangle, |1H2V+1H1V\rangle,
|1H2V+1H+1V\rangle$, $|1H1V+1H1V+1V\rangle, |1H1V+1H+2V\rangle$.

\noindent Their visibilities are listed in Table II.

\begin{table}
\caption{Visibility for 2 H-photons and 3 V-photons input }
\begin{ruledtabular}
\begin{tabular}{cccccccc}
\noalign{\smallskip}2H3V&2H2V&2H1V&1H3V&1H2V
&1H2V&HV+V&HV+V\\&+V&+2V&+H&+HV &+H+V&+HV& +H+V\\
\noalign{\smallskip}  \hline \noalign{\smallskip}
1&5/6&1/2&3/4&5/12&1/2&1/3&1/4 \smallskip
\end{tabular}
\end{ruledtabular}
\end{table}

\subsubsection{The Case of $|2_H, 4_V\rangle$}

The case of $|2_H, 4_V\rangle$ has 12 different scenarios. They
are

\noindent (i) $~|2H4V\rangle, |2H3V+1V\rangle, |2H2V+2V\rangle,
|2H1V+3V\rangle$, and

\noindent (ii) $|1H4V+1H\rangle, |1H3V+1H1V\rangle,
|1H3V+1H+1V\rangle,|1H2V+1H2V\rangle,$ $|1H2V+1H1V+1V\rangle,
|1H2V+1H+2V\rangle, |1H1V+1H1V+2V\rangle$, $|1H1V+1H+3V\rangle$.

The scenarios with different visibility are listed in Table III.
$|1H2V+1H1V+1V\rangle$ and $|1H2V+1H+2V\rangle$ have the same
visibility of 2/5 as $|1H2V+1H2V\rangle$.

\begin{table*}
\caption{Visibility for 2 H-photons and 4 V-photons input}
\begin{ruledtabular}
\begin{tabular}{cccccccccc}
 \noalign{\smallskip} 2H4V&2H3V&2H2V&2H1V
&1H4V&1H3V&1H3V&1H2V&2$\times$HV&1H1V\\&+V&+2V&+3V &+H&+HV&
+H+V&+1H2V&+2V&+1H+3V\\\noalign{\smallskip}
\hline\noalign{\smallskip}
 1&9/10&7/10&2/5&4/5&1/2&3/5&2/5&3/10&1/5 \\\noalign{\smallskip}
\end{tabular}
\end{ruledtabular}
\end{table*}

In general, they follow the trend that smaller visibility
corresponds to less photon overlapping. However, there are
exceptions: $1H2V+HV$ has less visibility than $1H2V+1H+V$ in
Table II and $1H3V+HV$ has less visibility than $1H3V+1H+1V$ in
Table III. So the runaway $HV$ does not help when $H$ and $V$
overlap in these cases.

\subsubsection{The Case of $|3_H, 3_V\rangle$}

There are totally 11 different scenarios in the special case of
$|3_H, 3_V\rangle$:

\noindent (i) $~|3H3V\rangle, |3H2V+V\rangle, |3H1V+2V\rangle$;

\noindent (ii) $|2H2V+1H1V\rangle, |2H2V+1H+1V\rangle,
|2H1V+1H2V\rangle$, $|2H1V+1H1V+1V\rangle, |2H1V+1H+2V\rangle$;

\noindent (iii) $|1H1V+1H1V+1H1V\rangle, |1H1V+1H1V+1H+1V\rangle$,
$|1H1V+1H+1V+1H+1V\rangle$.

In Table IV, we list the visibility for most of the scenarios.
$|2H1V+1H1V+1V\rangle$ and $|2H1V+1H+2V\rangle$ have the same
visibility of 2/5 as $|2H1V+1H2V\rangle$ and are not listed. As
can be seen, anomaly occurs for $|2H2V+1H1V\rangle$ and
$|2H2V+1H+1V\rangle$ where visibility is bigger for the case with
less photon overlap. The scenarios of $|3H3V\rangle$,
$|2H2V+1H1V\rangle$, and $|3\times HV\rangle$ were observed
experimentally by Xiang {\it et al.} \cite{xia} with the
corresponding visibility in Table IV.

\begin{table*}
\caption{Visibility for 3 H-photons and 3 V-photons input}
\begin{ruledtabular}
\begin{tabular}{ccccccccc}
\noalign{\smallskip} 3H3V&3H2V&3H1V&2H2V
&2H2V&2H1V&HV$\times$3&HV$\times$2&HV+V\\&+V&+2V&+HV&+H+V&+1H2V&
&+H+V&+H+H+V\\ \noalign{\smallskip}\hline \noalign{\smallskip}
 1 & 9/10 & 3/5&3/5&7/10&2/5&2/5&3/10&1/5 \\
\end{tabular}
\end{ruledtabular}
\end{table*}

\subsection{General Formula for the Visibility}

The most general case is when the input state is in the form of
$|k_H, N_V\rangle$ with $k\le N$. The most general scenario is
when the $k$ H-photons don't overlap in time but rather are split
into $r$ temporally well separated subgroups with $k_j$
indistinguishable photons in the $j$th group and $k_1+...+ k_r =
k$. We also divide the $N$ V-photons into $r+1$ subgroups with
$m_j$ V-photons overlap in time with the $j$th H-photon group. The
rest $N-m_1-...-m_r$ V-photons are in a separate group by
themselves. The wave function for these $N+k$ photons will satisfy
the permutation symmetry relation similar to Eq.(\ref{sym}) for
the overlapping photons and the orthogonal relation similar to
Eq.(\ref{orth}) for the well separated photons.

The derivation of the general formula for the visibility in the
$(N+k)$-photon NOON-state projection measurement is very
complicated and lengthy. It follows the general line of argument
as that leading to Eq.(\ref{VN+1}). We will present the detailed
procedure elsewhere \cite{ou3} but only give the result as
\begin{widetext}
\begin{eqnarray}
{\cal V}_{N+k} = \sum_{l=1}^k (-1)^{l-1}\sum_{i_1...i_r \atop
i_1+...+i_r=l}^{l} \bigg({l!\over i_1!...i_r!}\bigg)
{C_{k_1}^{i_1}...C_{k_r}^{i_r}~ m_1^{(i_1)}...m_r^{(i_r)}\over
(N+k-1)...(N+k-l)},~~~~~~~~\label{VN+k}
\end{eqnarray}
\end{widetext}
where $m^{(0)}=0=m^{(m)}$, $m^{(i)} \equiv m(m-1)...(m-i+1)$, and
$C_N^M\equiv (N+M)!/N!M!$. For the special case of $k=1$,
Eq.(\ref{VN+k}) recovers the expression in Eq.(\ref{VN+1}).
Furthermore, we can easily check that the formula in
Eq.(\ref{VN+k}) indeed leads to the visibility values in Tables
I-IV.

\section{Conclusion and Discussion}

The complementary principle of quantum interference is
demonstrated in a quantitative way in multi-photon interference
where photons can be categorized by their temporal
distinguishability. The temporal indistinguishability of photons
in turn can be characterized by the permutation symmetry in the
multi-photon wave function while the temporal distinguishability
by the orthogonality of the permuted wave functions.
Generalization to other degrees of freedom such as spatial modes
is straightforward. Although the above conclusions were made on
photons, they should apply to any bosons as well as fermions so
long as the occupation number of each mode is less than or equal
to one.

\begin{acknowledgments}
This work was supported by the US National Science Foundation
under Grant No. 0245421 and No.0427647. The author would like to
thank Mr. F. W. Sun for stimulating discussion.
\end{acknowledgments}

\end{document}